\documentstyle[11pt, epsf]{article}

\begin{document}

\title{Gamma-Ray Bursts from
\\Primordial Quark Objects in Space
\footnote{To be published in the
Proceedings of the Joint Meeting
of the Networks 'The Fundamental Structure
of Matter' and 'Tests of the Electroweak
Symmetry Breaking', Ouranoupolis, Greece,
May 1997.}
}
\author{B.~Anoushirvani, D.~Enstr\"{o}m, S.~Fredriksson, J.~Hansson, P.~\"{O}kvist
\\ Department of Physics \\
Lule\aa \ University of Technology\\
SE-971 87 Lule\aa , Sweden
\vspace{3mm}
\and A.~Nicolaidis
\\ Department of Theoretical Physics \\
Aristotle University of Thessaloniki \\
GR-540 06 Thessaloniki, Greece
\vspace{3mm}
\and S.~Ekelin
\\ Department of Mathematics \\
Royal Institute of Technology \\
SE-100 44 Stockholm, Sweden}

\date{ }

\maketitle

\begin{abstract}
We investigate the possibility that gamma-ray bursts
originate in a concentric spherical shell with a
given average redshift and find that this is indeed
compatible with the data from the third BATSE (3B)
catalog. It is also shown that there is enough freedom in
the choice of unknown burst properties to allow even for
extremely large distances to the majority of bursts.
Therefore, we speculate about an early,
and very energetic, origin of bursts, and suggest
that they come from phase transitions in massive
objects of pure quark matter, left over from the Big
Bang.
\end{abstract}

\newpage
{\flushleft The} bursts of intense gamma rays (GRB), first
observed in the 1960s by the Vela military satellites
designed to monitor breaches of the nuclear test ban
treaty of 1963, but disclosed to the civilian research
community only in 1973 \cite{Klebesadel},
have confounded physicists and astronomers ever
since. Although the outbursts must be very energetic,
the actual value of the total energy depends on
their distance from the earth. The time-span of the
bursts lie between around 0.01 and 1000
seconds, and no characteristic features, such as
spectral lines have been detected, with one exception.
The burst named GRB970508 has been related to
an object that appeared as an optical transient
shortly after the burst, revealing clear spectral
absorption lines. The absorbing body, which can be
either a host-galaxy of the GRB,
or an intervening foreground body, has been shown
to have a redshift parameter $z \approx 0.835$.
Hence the GRB source itself has $z \geq 0.835$.
There is also some indirect evidence for an
upper limit, $z \leq 2.3$ \cite{Metzger}.
There are still no clues to whether this GRB is
"average" in any sense, which means that detection
of future optical GRB transients with spectral
lines are certainly needed
before a distance-scale can be confirmed. Only
after such a scale has been determined, will
it be possible to discriminate between the many
dozen published theoretical models of the origin
of bursts. It should be noted that all other efforts
by authors of GRB publications to pinpoint an
absolute distance to a particular GRB, or a well-chosen
class of GRBs, are model-dependent, and therefore
less reliable. Such estimates seem to
cluster around $z$ values of $1 \div 2$.

In the 1980s, the consensus among researchers was
that the bursts originate within our own galaxy
\cite{grb-80}. When the Burst and Transient Source
Experiment (BATSE) \cite{BATSE}, aboard the
Compton Gamma Ray Observatory (CGRO), began
to produce much more data it became evident
that the gamma-ray bursts are distributed
isotropically in the sky, not following the visible
outlines of the Milky Way (nor of the Andromeda).
The opinion among astrophysicists then swayed
to models assuming a cosmological origin. A few
thousand gamma-ray bursts have been detected
to date, and there have been roughly as many
different publications on the subject.

The most popular GRB model seems to be that
they originate from the binary collapse of two very
compact star remnants; neutron stars, black holes,
or a combination thereof. Such models take it for
granted that these events occur at random in all
normal galaxies, typically once per a million years
per galaxy. The rarity of such mergers would explain
why none of the detected bursts has yet occurred close to
a visible galaxy, and why there has been no repetition
of events from the same locations. The low frequency
is also in line with estimates of the number of
neutron stars in galaxies, and even the energy release
seems to fit what would be expected if two neutron
stars merge. If the GRBs are evenly distributed
among galaxies, the bursts seem to release
$10^{51}-10^{53}$ erg of gamma rays.

Here we would like to test a completely different idea,
namely that GRBs are not at all evenly distributed in
space, with a universal frequency per galaxy, but
instead strongly biased toward large distances,
{\it i.e.}, the early Universe and high redshifts.
Since very distant GRBs must be more energetic
than in conventional models, we will also
speculate about their origin, although we will leave
the detailed work on a new model to a forthcoming
publication \cite{GRB2}.

Lacking an absolute distance-scale, it is, in fact,
almost trivial to fit a distribution of GRB distances,
with any chosen average distance, to the observations
of gamma-ray fluxes. We will demonstrate how this
works, with a simple choice of such a distribution.

We restrict ourselves to an
Einstein-de Sitter universe with vanishing
cosmological constant and global curvature.
This choice seems, by comparison to observational
data, to be a reasonably good approximation of the
Universe. We also assume that the individual bursts
can be treated as "standard candles", {\it i.e.},
that the characteristics of a typical ("average")
burst stays the
same during the full burst epoch.

Each burst is assumed to emit the radiation
uniformly in all directions ({\it i.e.}, not in
beams). Relaxing this condition would, of course,
require more bursts, and a lower energy
release per burst.

There seems to be no general agreement regarding
possible time-dilation effects in GRB spectra, nor
regarding an intrinsic duration-luminosity correlation
(incompatible with the standard candle assumption),
with strong bursts having shorter duration and
{\it vice versa}. We simply ignore such
(presumably weak) effects in the following analysis,
and concentrate on the number/peak-flux relation.

Taking one or more of these complications into account
would not change our general observation that a wide
range of GRB space distributions can be fitted
to the flux data.

The flux of a particular gamma-ray burst can, if the
conditions mentioned above are satisfied, be given as
a function of its redshift, $z$, \cite{MTW}

\begin{equation}
P(z) = \frac{L(z)}{4 \pi r(z)^2 (1 + z)^2},
\end{equation}

{\flushleft where} $L(z)$ is the luminosity of the
burst. The present distance to the source, $r(z)$,
depends on the cosmological model. In our case
(flat Einstein- de Sitter space), this relation reads

\begin{equation}
r(z) = \frac{2c}{H_0} (1 - \frac{1}{\sqrt{1 + z}}),
\end{equation}

{\flushleft where} $H_0$ is the Hubble constant
(taken as 75 km/s/Mpc).

The source luminosity detectable by an instrument
near the earth, with an effective energy detection
window between $E_{min}$ and $E_{max}$,
is given by

\begin{equation}
L(z) = \int_{E_{min} (1 + z)}^{E_{max}
(1 + z)} \phi (E) dE,
\end{equation}

{\flushleft where}

\begin{equation}
\phi (E) = A_0 \frac{e^{-E/kT}}{E}
\end{equation}

{\flushleft is} the spectral form (thermal
bremsstrahlung) conventionally chosen
for modelling the burst
\cite{Horack,Band,Bloom}. $kT$ is a characteristic
energy for a typical burst, chosen to be $350$~keV.
For BATSE, $E_{min} = 50$~keV and
$E_{max} = 300$~keV.

For simplicity, we assume that the number
density, $\rho (r)$, of the bursts is a gaussian,

\begin{equation}
\rho = C \, \frac{1}{\sqrt{2 \pi} \sigma}
e^{-(r - r_0)^2 / 2 \sigma^2},
\end{equation}

{\flushleft centred} around $r_0 = r(z_0)$,
and with variance $\sigma$. The normalising
constant $C$ is fitted to the data.
A homogeneous distribution in Euclidean
(fairly nearby) space, within a spherical shell
with nothing outside, is also compatible with 
the BATSE data within observational errors,
although such an abrupt cut-off seems unphysical.
A smoothed-out version of such a distribution,
or some completely different distribution altogether,
could equally well be fitted to the data. For brevity,
we only consider a gaussian distribution here.

In a given concentric spherical shell, there will
be a differential number of bursts given by

\begin{equation}
n = 4 \pi \, \rho \, r^2 .
\end{equation}

{\flushleft Equations} $(2)$ and $(5)$ then directly
give the parametric dependence, $n(z)$.

A typical fit to the BATSE data of $log(n) - log(P)$
is presented in Fig.~1, and the corresponding form
of $n$ as a function of $z$ in Fig.~2. It is
possible to fit the data reasonably well with any
choice of $z_0$, while $\sigma$ would be derived
by the fit (being smaller for higher $z_0$ value).

\begin{figure}
\centering
\epsfysize=6.5 cm
\leavevmode
\epsffile{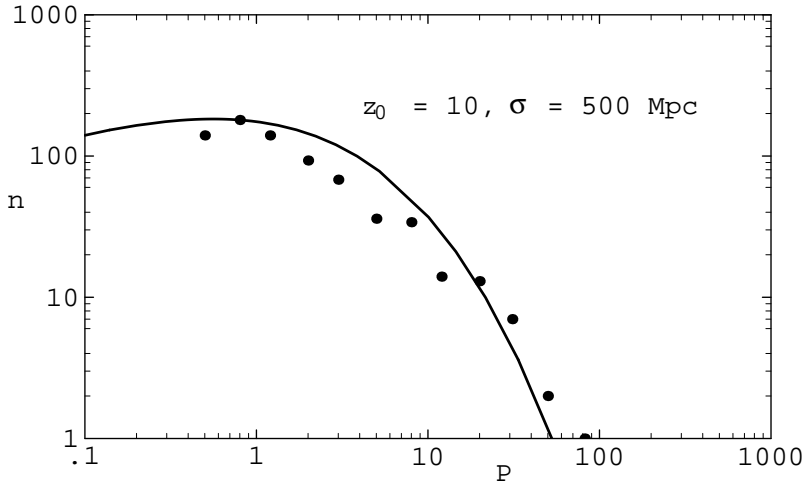}
\caption{A generic example of the differential
distribution of bursts ($n$), {\it versus} the peak
flux of registered photons ($P$) computed in
$256$ ms time-intervals. The data points
are from the BATSE 3B catalog, and the solid line
is the fit using a normal distribution peaked at
a redshift of $z_0 = 10$, with variance
$\sigma = 500$ Mpc.}
\label{n-p}
\end{figure}

\begin{figure}
\centering
\epsfysize=6.5 cm
\leavevmode
\epsffile{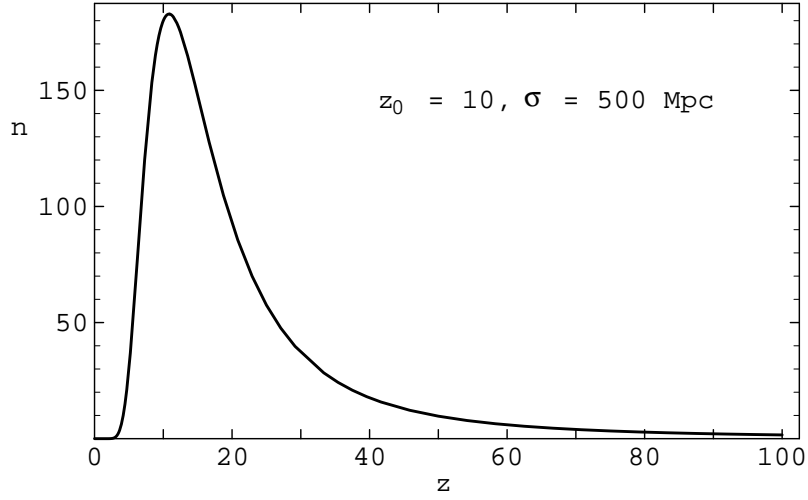}
\caption{The differential number of gamma-ray
bursts as a function of redshift in flat Einstein-de
Sitter space for a gaussian distribution with the
quoted values of $z_0$ and $\sigma$.}
\label{n-z}
\end{figure}

The data points are uncorrected for trigger
efficiency, as such a correction would overestimate
the true burst rate for fluxes near threshold (due
to not including atmospheric scattering), while the
data points with higher fluxes would be practically
unchanged.

\begin{figure}
\centering
\epsfysize=6.5 cm
\leavevmode
\epsffile{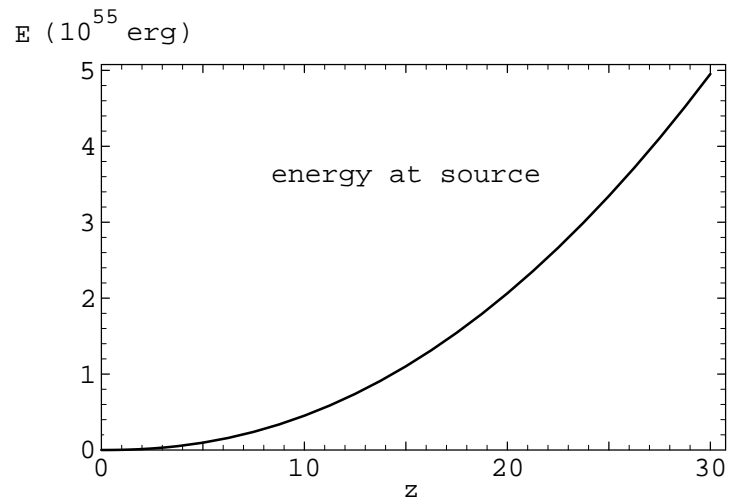}
\caption{The total gamma-ray energy emitted
by the source, given a flat Einstein-de Sitter space,
and a typical detected energy flow of
$10^{-5}$ erg/cm$^2$. The Hubble constant,
$H_0$, is taken as $75$ km/s/Mpc.}
\label{r-z}
\end{figure}

Suppose now that the bursts actually originate
at redshifts that, in the mean, are considerably
higher than the values $1 \div 2$, conventionally
discussed.
As can be seen from Fig.~3, standard GRBs
would then release more than $10^{54}$ erg of
energy (reaching, typically, $10^{59}$ erg at average
redshifts of $z_0 = 1000$).
We then have to consider mechanisms that would
radiate considerably more energy than
expected from, {\it e.g.}, neutron star mergers, and,
in addition, would be connected to very young
galaxies, or maybe even to the pregalactic era.
It is tempting to speculate that the bursts are
intimately related to the very creation of galaxies,
or of the normal visible matter in galaxies.
Since GRBs most probably come, directly or
indirectly, from detonations, we suggest that these
detonations are {\it phase transitions}
in large volumes of {\it quark matter} in the early
Universe.

In the conventional Big Bang scenario, the phase
transition from quark to normal nuclear matter
took place at a very early stage, as either a huge
detonation (second-order transition),
or via a multitude of rapidly growing hadronic
bubbles (first-order transition), once the global
temperature fell below some critical value. All
this occurred while the Universe was still one
enormous quark-gluon plasma.

In 1984 Witten \cite{Witten}, as well as
Fahri and Jaffe \cite{Fahri}, suggested that quark
matter might actually be the absolute ground state
of matter also at low temperatures, and that chunks
of such matter therefore could have escaped the
overall hadronisation after the Big Bang. Eventually,
these chunks could still be frequent enough to
make up the celebrated cosmic dark matter.

Strangely enough, this simple dark-matter candidate
has not gained the same popularity among
astro- and particle physicists as the
exotic ideas about neutralinos, axions,
heavy neutrinos and the like, or the more
down-to-earth brown dwarfs and "jupiters".
Maybe this is due to some rather involved and
model dependent counter arguments, like the
one in \cite{Madsen}, where it is claimed
that abundant quark "nuggets" should have
catalysed practically all neutrons stars into
quark stars, which, in turn, is claimed to be
inconsistent with some observational data.

Consequently, most theoretical work on
"quarks in space" now centres on the
possible creation of quark matter
out of normal matter in the cores of
extremely dense objects, such as neutron stars,
or collapsing supernovas. The engine for
this phase transition would then be external
gravitational pressure, which does not require
the pure quark matter to be the absolute ground
state of matter. Neither does it require any
virgin quark matter that has escaped the universal
phase transition.  There is, in fact, a model
built on the idea that the phase transition
{\it from} normal matter {\it to} quark
matter inside neutron stars is the true
source of gamma-ray bursts \cite{Cheng}.
An excellent review of the present understanding of
quark matter inside neutron stars has been
published recently by Glendenning \cite{Glend}.

Leaving the discussion of the dark-matter problem
to a forthcoming work \cite{GRB2}, we will now
discuss the possibility that a phase transition inside
leftover objects of pure quark matter is indeed
the source of gamma-ray bursts.

Our two crucial assumptions are that

(i) quark matter represents
the absolute ground state of "baryonic"
matter (at least above a certain mass),

(ii) the early, and rapidly
expanding, Universe
split up into such quark-matter objects
of different sizes {\it before} the
hadronisation into normal matter.

Hence, we would like to "revive" the original
idea by Witten, Fahri and Jaffe in a new version,
combined with a suggestion of a new ordering of
events after the Big Bang.

First, we note, however, that a sphere of
pure quark matter (being the ground-state
of matter or not) can be stable
only if its radius is less than around $14$ km.
A heavier object would simply collapse into a
black hole. Assume that the radius is
$R_{qm}$, the density is $\rho_{qm}$ and the
mass is $M_{qm}$. The condition for collapse is
then that the radius does not exceed the
Schwarzschild radius, given by

\begin{equation}
r_s = 2M_{qm}G/c^2,
\end{equation}

{\flushleft leading} to a critical radius

\begin{equation}
R_c = \sqrt {\frac{3c^2}{8 \pi \rho_{qm} G}}.
\end{equation}

{\flushleft Setting} $\rho_{qm} = \rho_{proton}$,
where the proton is assumed to have a radius
of $0.8$~fm, we get $R_c \approx 14$ km and
$M_c \approx 5M_{sun}$.

This means that the virgin quark-matter objects
with $R_{qm} \geq 14$ km could not
have survived for long. When their
internal expansion did no longer match the
overall expansion of the Universe, their cores
must have collapsed into black holes, leaving
the outer, and still expanding, layers relatively
intact. The exact moment for the creation of
the black hole is determined by the
balance between the speed of light,
the speed of sound (detonation wave) and
the speed of internal expansion.

Naturally, also lighter objects might be
unstable, and collapse after an initial
expansion, followed by a slower
gravitational contraction, in much the same
way as the development of a heavy star
into a supernova. We will analyse such a
slower collapse in a forthcoming work
\cite{GRB2}, where the balance between
gravitation and QCD forces will be
considered.

The quark-matter layers close to
the central black hole must,
however, have experienced a sudden,
enforced, drop of pressure and density,
which should have triggered a phase
transition to normal hadrons,
running from inside out. (This is {\it not} equivalent
to having the outermost layer of a quark-matter
sphere facing an empty surrounding.
An unlimited expansion outwards,
and a phase transition running from outside in,
is, in our model, prevented by gravity and strong
quark forces, {\it i.e.}, confinement.)

If the original quark-matter object was big enough,
it is likely that the phase transition, in the form of
a central detonation, did not embrace the whole object.
Rather, the outgoing particles (gammas, mesons,..),
expected from the hadronisation, should have ripped
the outer layers apart into smaller objects, before
they had time to experience the right
macroscopic conditions for a total phase transition.
This is again similar to a supernova detonation,
where the outermost parts are ripped apart by
neutrinos, while the interior is pushed into
a black hole or neutron star.

Such first-generation detonations most likely occurred
too early to be of interest as an explanation of GRBs.
The Universe was probably still too dense to let
out those gamma rays in such a virgin
shape that they can still be observed as pointlike
events. Typically, an exploding object giving rise
to the visible matter in a normal galaxy,
say the Milky Way, would have radiated
around $10^{64}$~erg of energy, assuming
that hadronisation gives 100~MeV of excess
energy per produced nucleon (give or take
little-known beaming effects due to
rotating plasmas, black-hole Kerr dynamics
and the like, being the origin,
or not, of disc-like galaxies).
Assuming a spherical quark object,
the inner part that hadronised should have had an
original radius of around $40,000$~km.

Nevertheless, each galaxy, or proto-galaxy, should
have a surrounding cloud of stable quark-matter
objects with radii less than $14$~km (or
whatever value a finer analysis will suggest).
Would they indeed make up the dark matter,
our galaxy would home at least $20$~billion
of them. Therefore, two such objects might
merge at any time, say after a random collision,
or after a slower spiralling within a binary
quark-matter system.
Would the new object become overcritical,
the chain of events, with a central black hole and
a detonating phase transition, would repeat.
The maximal mass of such an object
would then be twice the mass of a $14$~km sphere,
and the maximal release of excess energy would
be around $10^{54}$~erg, under the
idealised conditions that half of the mass goes into the
black hole and the other half hadronises
(radiating $100$~MeV per final nucleon).

In this way, a galaxy continues to create its own
nuclear matter through mergers of quark objects,
giving off gamma-ray bursts. We believe that this
process was more common in the distant
past, when quark matter was more abundant,
and when the merging of smaller proto-galaxies
might have been an important mechanism for
creating the galaxies we observe today.
An obvious advantage of such a scenario is
that we get gammas for free, without relying
on rare neutrino annihilation processes, and
subsequent interactions between charged
particles and the intergalactic medium.
Such processes are often claimed to be too
inefficient for producing the enormous
energy outbreaks that are observed.

{\it In conclusion}, we have shown that the gamma-ray
bursts might well come from very distant sources.
We suggest that these sources are connected to
phase transitions of big objects of pure quark
matter into normal nuclear matter, taking place
in young galaxies, during the merging of smaller
proto-galaxies, or maybe even at the very moment
of the creation of a whole galaxy.
The phase transitions are triggered
by black holes, formed when two
quark-matter objects merge into an
overcritical system, or inside an object
that was overcritical all from the Big Bang.
If so, the gamma-ray bursts are indeed the
ultimate cosmic
fireworks, announcing the birth of matter
as we know it!

One of us (S.F.) would like to thank the
organisers of this meeting for kind
hospitality and for creating a most inspiring
atmosphere. This project is supported by the
European Commission under contract
CHRX-CT94-0450, within the network
"The Fundamental Structure of Matter".


\begin{thebibliography}{45}
\bibitem{Klebesadel} R. W. Klebesadel, I. B. Strong,
R. A. Olson, Ap. J. Lett. {\bf 82}, L85 (1973).
\bibitem{Metzger} M. R. Metzger {\it et al.},
Nature {\bf 387}, 878 (1997).
\bibitem{grb-80} A. K. Harding, Phys. Rep.
{\bf 206}, 327 (1991).
\bibitem{BATSE} C. A. Meegan {\it et al.},
Ap. J. Suppl. {\bf 106}, 65 (1996).
\bibitem{GRB2} D. Enstr\"{o}m {\it et al.},
in preparation, and D. Enstr\"{o}m,
Lule\aa \ University MSc diploma thesis,
in print.
\bibitem{MTW} See any standard text, {\it e.g.},
C. W. Misner, K. S. Thorne, J. A. Wheeler,
{\it Gravitation}, Freeman 1973; or S. Weinberg,
{\it Gravitation and Cosmology}, Wiley 1972.
\bibitem{Horack} J. M. Horack, R. S. Malozzi,
T. M. Koshut, Ap. J. {\bf 466}, 21 (1996).
\bibitem{Band} D. L. Band {\it et al.},
Ap. J. {\bf 413}, 281 (1993).
\bibitem{Bloom} E. E. Fenimore, J. S. Bloom,
Ap. J. {\bf 453}, 25 (1995).
\bibitem{Witten} E. Witten, Phys. Rev
{\bf D30}, 272 (1984).
\bibitem{Fahri} E. Fahri, R. L. Jaffe,
Phys. Rev. {\bf D30}, 2379 (1984).
\bibitem{Madsen} J. Madsen, Phys. Rev.
Lett. {\bf 61}, 2909 (1988).
\bibitem{Cheng} K. S. Cheng, Z. G. Dai,
Phys. Rev. Lett. {\bf 77}, 1210 (1996).
\bibitem{Glend} N. K. Glendenning,
{\it Compact Stars}, Springer-Verlag 1996.



\end{thebibliography}
\end{document}